\documentclass[prd,amsmath,amssymb,twocolumn]{revtex4-1}

\usepackage{graphicx}
\usepackage{color}
\usepackage{amsmath}
\usepackage{amssymb}

\usepackage{bm}% bold math
\usepackage{bbm}
\usepackage{MnSymbol}

\newcommand{\RCHO}{\mathbb{R}\otimes\mathbb{C}\otimes\mathbb{H}\otimes\mathbb{O}}

\newcommand{\CLten}{\mathbb{C}l(10)}

\newcommand{\C}{\mathbb{C}}
\newcommand{\R}{\mathbb{R}}

\usepackage{xcolor}

\usepackage{color}
 
\begin{document}

\title{An Algebraic Roadmap of Particle Theories  \vspace{2mm}\\ \it Part II:  Theoretical checkpoints\rm}

\author{N. Furey}
\affiliation{$ $\\  Iris Adlershof, Humboldt-Universit\"{a}t zu Berlin,\\ Zum Grossen Windkanal 2, Berlin, 12489 \vspace{2mm} \\ furey@physik.hu-berlin.de \\HU-EP-23/65 }

\begin{abstract}

An optimal algebraic model of particle physics has a number of checkpoints to pass.  As a minimum, models should $\langle 1 \rangle$~conform to the Coleman-Mandula theorem (or establish a loophole), $\langle 2 \rangle$~evade familiar fermion doubling problems, $\langle 3 \rangle$~naturally explain the Standard Model's chirality, $\langle 4 \rangle$~exclude B-L gauge symmetry at low energy, and $\langle 5 \rangle$~explain the existence of three generations.  We  demonstrate how the model introduced in~\citep{fr1} passes checkpoints $\langle 1 \rangle, \langle 2 \rangle, \langle 3 \rangle, \langle 4 \rangle$, and has yet to cross~$\langle 5 \rangle$.  We close by elucidating an unexpected appearance of spacetime symmetries.
 \end{abstract}

\maketitle

This article relies on the results and conventions of~\citep{fr1}.

\section{Five checkpoints\label{miles}}

The last decades have seen a significant number researchers steadily unearth more and more  algebraic structure underlying the Standard Model of particle physics, \citep{fr1}-\citep{osmu-furey}.  With recent meetings, \citep{PI}, \citep{APQ}, \citep{OSMU}, this community continues to grow.

Needless to say, however, a full solution is still at large.  Certain checkpoints must necessarily be crossed before one can finally claim a compelling algebraic Standard Model description.  We find that five challenges in particular recur in the literature again and again (including in our own models).  We search for a concise algebraic explanation for the Standard Model's structure that overcomes the following hurdles:

$\langle 1 \rangle$  \bf Heed or evade the Coleman-Mandula theorem.  \rm Internal and  spacetime symmetries should either commute on particle representations, or a loophole should be established.

$\langle 2 \rangle$ \bf Solve the fermion doubling problem.  \rm The Hilbert space corresponding to one generation of standard model fermions is locally 32$\hspace{.5mm}\C$ dimensional.  Likewise, under $\mathfrak{spin}(10) \oplus \mathfrak{spin}(1,3),$ unconstrained $(\mathbf{16}, \mathbf{2})$ Weyl  representations corresponding to one generation of fermions is also locally 32$\hspace{.5mm}\C$ dimensional.  
%(It is common, and problematic, to double a 32$\hspace{.5mm}\C$ dimensional spinor space in order to introduce $\mathfrak{sl}(2,\C)$ spacetime transformations).

$\langle 3 \rangle$  \bf Explain the Standard Model's chirality.  \rm  Here we mean without implementing \it ad hoc \rm projection operators, without fixing arbitrarily chosen mathematical objects, without  introducing other  \it ad hoc \rm constraints.

$\langle 4 \rangle$  \bf Exclude the extraneous \rm $\mathfrak{u}(1)_{B-L}$ \bf symmetry at low energy. \rm  This additional gauge symmetry often appears in algebraic models.  Again, it should be eliminated on logical grounds, not explained away with an arbitrarily chosen condition.

$\langle 5 \rangle$  \bf  Explain the existence of three generations. \rm   One would anticipate that these three generations should be linearly independent from one another, or that an explanation should be provided as to why their linear dependence is desired.

This list extends the one originally presented in~\citep{fh1}.  In our recent work, \citep{fh1}, \citep{fh2}, we demonstrated solutions for $\langle 1 \rangle$ and $\langle 2 \rangle$, but not $\langle 3 \rangle$, $\langle 4 \rangle$, $\langle 5 \rangle$.  In earlier work, \citep{malala}, it could be argued that we achieved solutions for $\langle 1 \rangle$ and $\langle 3 \rangle$, but not $\langle 2 \rangle$, $\langle 4 \rangle$, $\langle 5 \rangle$.

\begin{figure}[h!]
\begin{center}
\includegraphics[width=9.6cm]{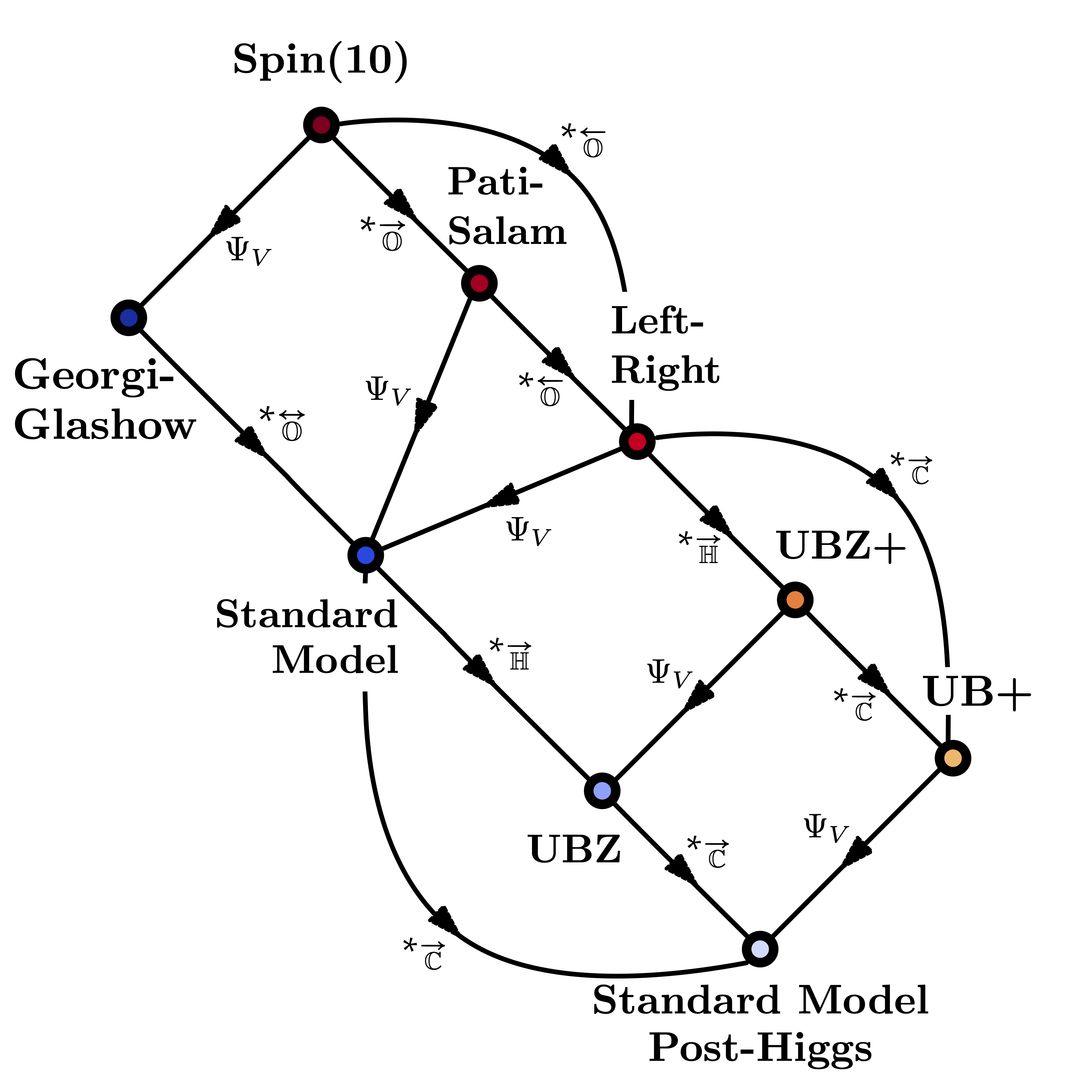}
\caption{\label{5step}
A  particle roadmap described in~\citep{fr1} interconnecting  six well-known models:  Spin(10), Pati-Salam (PS), Left-Right Symmetric (LR), Georgi-Glashow (SU(5)), Standard Model pre-Higgs-mechanism (SM), Standard Model post-Higgs-mechanism (UB).}
\end{center}
\end{figure}

The purpose of this current article is to work through the implications of our general construction in~\citep{fr1}.  We begin by  reminding the reader of the detailed particle roadmap, first constructed in~\citep{fr1}, depicted here in Figure~(\ref{5step}).  The majority of this article is then devoted to demonstrating that our new model passes checkpoints $\langle 1 \rangle$, $\langle 2 \rangle$, $\langle 3 \rangle$, $\langle 4 \rangle$, and has yet to overcome $\langle 5 \rangle$.  In this context, we finally show how SL(2,$\C$) spacetime symmetries arise readily from the maximal Clifford subspace commuting with Spin(10).

\section{An early checkpoint}

In our previous paper,~\citep{fr1}, we showed that it is possible to use the multivector condition, $\Psi_V$, in order to arrive at the Lie algebra for the Standard Model, $\ell_{\textup{SM}}$,  
\begin{equation}\begin{array}{rccll}\label{lsm}
\ell_{\textup{SM}}\Psi_{\textup{L}} = \big( r_j i \Lambda_j &+& r'_m L_{\epsilon_m}s  &+& irY\big)\Psi_{\textup{L}}.\vspace{2mm}\\
\mathfrak{su}(3)&\oplus&\mathfrak{su}(2)&\oplus&\mathfrak{u}(1)
\end{array}\end{equation}
\noindent Unlike in our earlier work,~\citep{fh2}, this result now excludes an unwanted B-L gauge symmetry at low energy.  In~\citep{fr1}, we defined the B-L operator as
\begin{equation}i(B-L)\Psi_{\textup{L}} := \frac{1}{3}(L_{e_1}L_{e_3}+L_{e_2}L_{e_6}+L_{e_4}L_{e_5})\Psi_{\textup{L}}.
\end{equation}
Given that the multivector condition may be thought of as a consistency condition when spinors are constructed from multivectors, this qualifies as a natural explanation for the omission of B-L.  Hence, we see that this model  passes checkpoint $\langle 4 \rangle$.

\section{Fermion identification}

\subsection{The even semi-spinor}

We may now apply the Standard Model symmetries above to $\Psi_{\textup{L}}$ so as to label its states, and confirm that the electroweak sector is maximally chiral (checkpoint $\langle 3 \rangle$).  We label these below, according to their behaviour under $\ell_{\textup{SM}}$ of equation~(\ref{lsm}).  The B-L operator is also useful in this task, as it differentiates between sterile neutrinos and sterile antineutrinos, where the Standard Model symmetries do not.
\begin{equation}\begin{array}{l}\label{PsiL_ladder}
\Psi_{\textup{L}}  \hspace{1mm}=\hspace{1mm}\frac{1}{2}\left(1+i\Gamma_{11}\right)\CLten \hspace{.5mm} v \vspace{3mm}\\

 =  \hspace{.5mm} {\mathcal{V}}_L \hspace{.5mm} a_4a_3a_2a_1\hspace{.5mm} v + \mathcal{E}^{-}_L \hspace{.5mm} a_5a_3a_2a_1\hspace{.5mm} v  + \overline{\mathcal{D}^{ k}_R} \hspace{.5mm}\epsilon_{ijk}a_5a_4a_ja_i\hspace{.5mm} v  \vspace{3mm}\\

 + \hspace{1mm} \overline{\mathcal{U}_R^{  k} }\hspace{.5mm} \epsilon_{ijk}a_ja_i\hspace{.5mm} v +    {\mathcal{U}}_L^{  i} \hspace{.5mm} a_4 a_i \hspace{.5mm} v  +    {\mathcal{D}}_L^{  i} \hspace{.5mm} a_5  a_i \hspace{.5mm} v  +    \overline{{\mathcal{E}}^-_R} \hspace{1mm} a_5  a_4 \hspace{.5mm} v \vspace{3mm}\\

+ \hspace{1mm}  \overline{\mathcal{V}_R} \hspace{.5mm}\hspace{.5mm} v,
\end{array}
\end{equation}
\noindent where $i,j,k\in\{1,2,3\}$.  Here, the coefficients $\mathcal{V}_L,$ $\mathcal{E}^-_L$, $\dots$ are suggestively named complex coefficients.  For example, $\mathcal{V}_L$ refers to the left-handed neutrino,  $\mathcal{E}^{-}_L$ refers to the left-handed electron, $\overline{\mathcal{D}^{ k}_R}$ refers to the anti-particle of the right-handed down quark (which is itself left-handed), etc.  This convention is chosen so as to coincide with~\citep{fh1}, \citep{fh2}, but differs from~\citep{AGUTS}, \citep{thesis}.

Substituting in the division algebraic realizations of these states via equations~(25) and (28) of our previous paper,~\citep{fr1}, gives $\Psi_{\textup{L}}$ as 
\begin{equation}\begin{array}{lll}\label{PsiL_div}
\Psi_{\textup{L}}  \hspace{1mm}&=&  \hspace{.5mm} \Large(\hspace{.5mm} {\mathcal{V}}_L \hspace{.5mm}  i\alpha_1 \alpha_2 \alpha_3 L_{\epsilon_{\uparrow\downarrow}}R_{\epsilon_{\downarrow\downarrow}}  \vspace{3mm}\\

 &+& \mathcal{E}^{-}_L \hspace{.5mm} i\alpha_3 \alpha_2 \alpha_1 L_{\epsilon_{\downarrow\downarrow}}R_{\epsilon_{\downarrow\downarrow}}  + \overline{\mathcal{D}^{ k}_R} \hspace{.5mm}\epsilon_{ijk}\alpha_j \alpha_i L_{\epsilon_{\uparrow\downarrow}}R_{\epsilon_{\downarrow\downarrow}}   \vspace{3mm}\\

 &+& \hspace{1mm} \overline{\mathcal{U}_R^{  k} }\hspace{.5mm} \epsilon_{ijk}\alpha_j\alpha_iL_{\epsilon_{\downarrow\downarrow}}R_{\epsilon_{\downarrow\downarrow}} -    {\mathcal{U}}_L^{  i} \hspace{.5mm}  i\alpha_i  L_{\epsilon_{\uparrow\downarrow}}R_{\epsilon_{\downarrow\downarrow}}     \vspace{3mm}\\
 
&+&   {\mathcal{D}}_L^{  i} \hspace{.5mm} i\alpha_i  L_{\epsilon_{\downarrow\downarrow}}R_{\epsilon_{\downarrow\downarrow}} +    \overline{{\mathcal{E}}^-_R} \hspace{1mm} L_{\epsilon_{\uparrow\downarrow}}R_{\epsilon_{\downarrow\downarrow}} \vspace{3mm}\\

&+& \hspace{1mm}  \overline{\mathcal{V}_R} \hspace{.5mm}\hspace{.5mm} L_{\epsilon_{\downarrow\downarrow}}R_{\epsilon_{\downarrow\downarrow}}\Large)\hspace{1mm}s^*S^*,
\end{array}
\end{equation}
\noindent where $i,j,k\in\{1,2,3\},$ 
\begin{equation}\begin{array}{ll}
s:=\frac{1}{2}\left(1+iL_{e_7}\right), \hspace{1cm}&S:=\frac{1}{2}\left(1+iR_{e_7}\right), \vspace{2mm}\\
s^*:=\frac{1}{2}\left(1-iL_{e_7}\right), &S^*:=\frac{1}{2}\left(1-iR_{e_7}\right), \vspace{2mm}\\
\epsilon_{\uparrow\uparrow}:= \frac{1}{2} \left(1 +i\epsilon_3\right), &\epsilon_{\downarrow\uparrow}:= \frac{1}{2} \left(\epsilon_2 +i\epsilon_1\right),\vspace{2mm}\\
\epsilon_{\uparrow\downarrow}:= \frac{1}{2} \left(-\epsilon_2 +i\epsilon_1\right), &\epsilon_{\downarrow\downarrow}:= \frac{1}{2} \left(1 -i\epsilon_3\right),
\end{array}\end{equation}
\noindent and the $\alpha_i$ are defined so as to match previous articles, \citep{fh1}, \citep{fh2}, \citep{thesis}, \citep{Gen}, \citep{321}.  Namely, 
\begin{equation}\begin{array}{l} \alpha_1 := \frac{1}{2}\left(-L_{e_5}+iL_{e_4}\right),\vspace{2mm}\\ 
\alpha_2 := \frac{1}{2}\left(-L_{e_3}+iL_{e_1}\right), \vspace{2mm}\\
\alpha_3 := \frac{1}{2}\left(-L_{e_6}+iL_{e_2}\right).
\end{array}\end{equation}

\subsection{The odd semi-spinor}

It is clear from the $\mathbb{Z}_2$-grading inherent in the minimal left ideal, $\Psi,$ that these left-handed particles, $\Psi_{\textup{L}},$ comprise the even-graded subspace.   How might we then interpret $\Psi$'s odd-graded subspace?  

It will be of little surprise to the reader that when we choose  $\mathfrak{spin}(10)$ acting on $\Psi$ to be represented by linear combinations of Clifford algebraic bivectors, $\Gamma_i \Gamma_j,$ then this 32$\hspace{.5mm}\C$ dimensional $\Psi$ decomposes as the $\mathbf{16}\oplus \mathbf{16^*}$ of $\mathfrak{spin}(10).$  In terms of division algebras,  these $\mathfrak{spin}(10)$ Lie algebra elements may be encoded as
\begin{equation}\begin{array}{ll}\label{so10D}
\ell_{10}^D:=&r_{ij}L_{e_i}L_{e_j} + r_m L_{\epsilon_m} + r'_m L_{\epsilon_m}L_{e_7}R_{\epsilon_3} \vspace{2mm}
\\&+\hspace{1mm} r_{mj}' iL_{\epsilon_m}L_{e_j} + r_j''iL_{e_j}L_{e_7}R_{\epsilon_3},
\end{array}\end{equation}
\noindent for $i,j\in\{1,\dots 6\}$ with $i\neq j,$  $m\in\{1,2, 3\},$ and $r_{ij}, r_m, r'_m, r_{mj}', r_j''\in\R$.  In this case, the odd-graded space is interpreted as a right-handed semi-spinor:
\begin{equation}\begin{array}{ll}\label{PsiR}
 \Psi_{\textup{R}} \hspace{1mm}&=\hspace{1mm}\frac{1}{2}\left(1-i\Gamma_{11}\right)\CLten \hspace{.5mm} v \vspace{3mm}\\

 &= \mathcal{V}_R^{ } \hspace{.5mm}a_5 a_4 a_3 a_2 a_1 \hspace{.5mm} v   \vspace{3mm}\\

&+ \hspace{.5mm}\mathcal{E}^{  -}_R \hspace{.5mm}a_3 a_2 a_1 \hspace{.5mm} v  +  \overline{\mathcal{D}_L^{  k}} \hspace{1mm} \epsilon_{ijk}a_4 a_j a_i \hspace{.5mm} v  \vspace{3mm}\\

&+ \hspace{.5mm} \overline{\mathcal{U}_L^{  k}} \hspace{1mm} \epsilon_{ijk}a_5 a_j a_i \hspace{.5mm} v  +  {\mathcal{U}_R^{  i}} \hspace{.5mm} a_5 a_4 a_i \hspace{.5mm} v \vspace{3mm}\\

&+ \hspace{.5mm} \mathcal{D}_R^{i} \hspace{.5mm} a_i^{\dagger}\hspace{.5mm} v   + \hspace{.5mm}\overline{\mathcal{E}^{  -}_L} \hspace{.5mm}a_4\hspace{.5mm} v    +  \overline{\mathcal{V}_L} \hspace{.5mm}a_5 \hspace{.5mm} v.  
\end{array}
\end{equation}
\noindent Again, we have made use of the Standard Model + B-L Lie subalgebra in order to identify states.    In terms of division algebras, $\Psi_{\textup{R}}$ becomes
\begin{equation}\begin{array}{lll}\label{PsiR_div}
\Psi_{\textup{R}}  \hspace{1mm}&=&  \hspace{.5mm} \Large(\hspace{.5mm} {\mathcal{V}}_R \hspace{.5mm}  i\alpha_3 \alpha_2 \alpha_1 L_{\epsilon_{\uparrow\downarrow}}R_{\epsilon_{\downarrow\uparrow}}  \vspace{3mm}\\

 &+& \mathcal{E}^{-}_R \hspace{.5mm} i\alpha_3 \alpha_2 \alpha_1 L_{\epsilon_{\downarrow\downarrow}}R_{\epsilon_{\downarrow\uparrow}}  + \overline{\mathcal{D}^{ k}_L} \hspace{.5mm}\epsilon_{ijk}\alpha_j \alpha_i L_{\epsilon_{\uparrow\downarrow}}R_{\epsilon_{\downarrow\uparrow}}   \vspace{3mm}\\

 &-&  \overline{\mathcal{U}_L^{  k} }\hspace{.5mm} \epsilon_{ijk}\alpha_j\alpha_iL_{\epsilon_{\downarrow\downarrow}}R_{\epsilon_{\downarrow\uparrow}} +    {\mathcal{U}}_R^{  i} \hspace{.5mm}  i\alpha_i  L_{\epsilon_{\uparrow\downarrow}}R_{\epsilon_{\downarrow\uparrow}}     \vspace{3mm}\\
 
&+&   {\mathcal{D}}_R^{  i} \hspace{.5mm} i\alpha_i  L_{\epsilon_{\downarrow\downarrow}}R_{\epsilon_{\downarrow\uparrow}} +    \overline{{\mathcal{E}}^-_L} \hspace{1mm} L_{\epsilon_{\uparrow\downarrow}}R_{\epsilon_{\downarrow\uparrow}} \vspace{3mm}\\

&-&  \overline{\mathcal{V}_L} \hspace{.5mm}\hspace{.5mm} L_{\epsilon_{\downarrow\downarrow}}R_{\epsilon_{\downarrow\uparrow}}\Large)\hspace{1mm}s^*S^*.
\end{array}
\end{equation}

With this identification of $\Psi_{\textup{R}}$, the full minimal left ideal, $\Psi = \Psi_{\textup{L}}\oplus \Psi_{\textup{R}},$ transforms reducibly as $\mathbf{32} = \mathbf{16}\oplus  \mathbf{16^*}$ under $\mathfrak{spin}(10)$.  It is therefore appropriate for use as a 32$\hspace{.5mm}\C$ dimensional Hilbert space for the Standard Model at high energies.  This matches the representation space described in~\citep{AGUTS} and \citep{thesis}.

Readers may confirm through application of equation~(\ref{lsm}) and its right-handed counterpart, that this new model is indeed chiral, and hence passes checkpoint $\langle 3 \rangle$.

Finally, we note for future reference that the maximal Lie subalgebra in $\CLten$ commuting with elements $\ell_{10}^D$  is generated by two objects over $\C$, namely $\{1, \Gamma_{11} \}$.
\begin{equation}\label{max} [\hspace{.5mm}\ell_{max}, \hspace{1mm}\ell_{10}^D\hspace{.5mm}]=0 \hspace{4mm}\Rightarrow\hspace{4mm} \ell_{max} = c_0 + c_1 \Gamma_{11}
\end{equation}
\noindent for $c_0, c_1 \in \C$.

\section{Spacetime symmetries as the complement of internal symmetries}

Readers should note the unusual way in which local spacetime symmetries arise in this model. 

\subsection{The last division algebraic reflection}

It can be seen that the Dirac spinor $\mathbf{32}=\mathbf{16}\oplus  \mathbf{16^*}$ is well-suited so as to describe a fermionic Hilbert space, \citep{AGUTS}, \citep{thesis}. However, these representations do not match those of the unconstrained fields one would typically use to build a Lagrangian.  Under local symmetries $\left(\mathfrak{spin}(10), \hspace{1mm} \mathfrak{sl}(2,\C)\right)$ the representation $\mathbf{32}=\left(\mathbf{16}, \mathbf{2} \right)$ would be more appropriate, where $\mathfrak{sl}(2,\C)$ enacts local spacetime symmetries, and our fermions are represented as $2\hspace{.5mm}\C$ dimensional Weyl representations.

How does one bring about this $\left(\mathbf{16}, \mathbf{2} \right)$ representation?  It is important to note that the Coleman-Mandula theorem and equation~(\ref{max}) imply that the $\ell_{10}^D$ of equation~(\ref{so10D}) do not provide the $\mathfrak{spin}(10)$ action that we need.  Instead, we will need a $\mathfrak{spin}(10)$ action that fashions the 32 $\C$ dimensional minimal left ideal into a $\mathbf{16}\oplus  \mathbf{16}.$  In~\citep{fh1} such an action was found, however, at the time we did not explain how the solution was originally obtained.  

%Again, our idea to implement Weyl spinors in order to introduce spacetime symmetries was discussed repeatedly with colleagues, and subsequently implemented into their models without attribution.  It was this finding that solved the fermion doubling problem,~\citep{fh1}.

In \citep{fr1}, we made use of four division algebraic reflections:  ${*_{\overrightarrow{\mathbb{O}}}}$,  $*_{\overleftarrow{\mathbb{O}}}$, $*_{\overrightarrow{\mathbb{H}}}$, $*_{\overrightarrow{\mathbb{C}}}$ in relation to the internal symmetries of our model.  Readers may notice that these four types of reflection correspond to the four types of multiplication algebra that are expressible in terms of \it left \rm multiplication.  (Please refer to Sections IV D and IV F of~\citep{fr1}).  Beyond these four types of reflection, only one remained, $*_{\overleftarrow{\mathbb{H}}}$, and its corresponding multiplication algebra is \it not \rm expressible in terms of left multiplication.

For concreteness, let us take $*_{\overleftarrow{\mathbb{H}}}\vspace{.5mm}$, to define the map $R_{\epsilon_1}\mapsto R_{\epsilon_1},$ $R_{\epsilon_2}\mapsto -R_{\epsilon_2},$ $R_{\epsilon_3}\mapsto -R_{\epsilon_3}.$  With this final involution at our disposal, we may generalize the Majorana action defined in Section 3.5 of \citep{thesis}.  That is, instead of relying on $\ell_{10}^D$ of equation~(\ref{so10D}), we will consider  elements of $\mathfrak{spin}(10)$ represented as
\begin{equation}\begin{array}{ll}  &\ell_{10}^D \hspace{1mm}P\hspace{1mm} + \hspace{1mm}\left({\ell_{10}^D}\hspace{1mm} P\right)^{*_{\overleftarrow{\mathbb{H}}}}  \vspace{2mm} \\

= &\ell_{10}^D \hspace{1mm}P\hspace{1mm} + \hspace{1mm}{\ell_{10}^D}^{*_{\overleftarrow{\mathbb{H}}}}\hspace{1mm} P^* \vspace{2mm}\\

= &  r_{ab}L_{e_a}L_{e_b} + r'_{mn} L_{\epsilon_m}L_{\epsilon_n} + r_{ma}'' iL_{\epsilon_m}L_{e_a} \vspace{2mm}\\

= & \ell_{10},
\end{array}\end{equation}
\noindent familiar from equation~(34) of reference~\citep{fr1}.  Here, the idempotent $P$ is defined as $P:=\frac{1}{2}\left(1+i\Gamma_{11}\right).$

When we apply this $*_{\overleftarrow{\mathbb{H}}}\vspace{.5mm}$-invariant action to the entire minimal left ideal, $\Psi = \CLten \hspace{.5mm} v$, then this 32 $\C$ dimensional space  transfoms as a $\mathbf{16}\oplus  \mathbf{16}$, as desired.

\subsection{An unexpected appearance of spacetime symmetries}

% Show how it relates to Dynkin diagrams.

With this new $*_{\overleftarrow{\mathbb{H}}}\vspace{.5mm}$-invariant $\mathfrak{spin}(10)$ action, we may now ask again:  What is the maximal Lie subalgebra in $\CLten$ commuting with it?  In contrast to equation~(\ref{max}), we now find that 
\begin{equation}\label{max2} [\hspace{.5mm}\ell_{max}, \hspace{1mm}\ell_{10}\hspace{.5mm}]=0 \hspace{4mm}\Rightarrow\hspace{4mm} \ell_{max} = c_{\mu}R_{\epsilon_{\mu}}
\end{equation}
\noindent for $\mu \in \{0,1,2,3\},$ and  $c_{\mu} \in \C$.

In other words, \it the maximal space commuting with this version of $\mathfrak{spin}(10)$ is given by the trivial centre of $\CLten$, and  $\mathfrak{sl}(2,\C)$.  \rm Furthermore, this $\mathfrak{sl}(2,\C)$ action behaves precisely as do local spacetime symmetries on Weyl representations.  It is perhaps intriguing that the maximal subspace commuting with $\mathfrak{spin}(10)$ should almost exclusively act as a source for spacetime symmetries.  

Schematically, $*_{\overleftarrow{\mathbb{H}}}$ has allowed us to redefine
%\begin{equation}\begin{array}{lll}
%\delta \Psi = \ell_{10}^D\hspace{.5mm}\Psi &\hspace{2.5mm}\mapsto\hspace{2.5mm}&\delta \Psi = \ell_{10}^D\hspace{.5mm}P\hspace{.5mm}\Psi +(\ell_{10}^{D}\hspace{.5mm}P\hspace{.5mm}\Psi )^{*_{\overleftarrow{\mathbb{H}}}} 
%\vspace{2mm}\\&&\hspace{.53cm}= \ell_{10}\hspace{.5mm}\Psi\vspace{2mm}\\
%\Psi\sim\mathbf{16}\oplus\mathbf{16^*}&\hspace{2.5mm}\mapsto\hspace{2.5mm}&\Psi\sim \mathbf{16}\oplus\mathbf{16}\vspace{2mm}\\
%\ell_{\textup{max}} = \C\oplus\C &\hspace{2.5mm}\mapsto\hspace{2.5mm}&\ell_{\textup{max}} = \C\oplus sl(2,\C),
%\end{array}\end{equation}
\begin{equation}\begin{array}{lll}
\delta \Psi = \ell_{10}^D\hspace{.5mm}\Psi &\hspace{2.5mm}\mapsto\hspace{2.5mm}&\delta \Psi = \ell_{10}^D\hspace{.5mm}P\hspace{.5mm}\Psi +(\ell_{10}^{D}\hspace{.5mm}P)^{*_{\overleftarrow{\mathbb{H}}}} \hspace{.5mm}\Psi  = \ell_{10}\hspace{.5mm}\Psi\vspace{2mm}\\
\Psi\sim\mathbf{16}\oplus\mathbf{16^*}&\hspace{2.5mm}\mapsto\hspace{2.5mm}&\Psi\sim \mathbf{16}\oplus\mathbf{16}\vspace{2mm}\\
\ell_{\textup{max}} \in \C\oplus\C &\hspace{2.5mm}\mapsto\hspace{2.5mm}&\ell_{\textup{max}} \in \C\oplus \mathfrak{sl}(2,\C).
\end{array}\end{equation}

The new symmetry actions permit us to compile and re-identify our fermionic states as
\begin{equation}\begin{array}{l}\label{S32}
\Psi_{\textup{L}}^{\uparrow} + \Psi_{\textup{L}}^{\downarrow} \hspace{1mm}=\hspace{1mm}\CLten \hspace{.5mm} v \vspace{3mm}\\

 = \mathcal{V}_L^{\uparrow } \hspace{.5mm}a_5  a_4  a_3  a_2  a_1  \hspace{.5mm} v  + {\mathcal{V}}_L^{\downarrow} \hspace{.5mm} a_4  a_3  a_2  a_1  \hspace{.5mm} v \vspace{3mm}\\

 + \hspace{.5mm} \mathcal{E}^{-\downarrow}_L \hspace{.5mm} a_5  a_3  a_2  a_1  \hspace{.5mm} v  + \overline{\mathcal{D}^{ k\uparrow}_R} \hspace{.5mm}\epsilon_{ijk}a_5  a_4  a_j  a_i  \hspace{.5mm} v  \vspace{3mm}\\

+ \hspace{.5mm}\mathcal{E}^{  -\uparrow}_L \hspace{.5mm}a_3  a_2  a_1  \hspace{.5mm} v  +  \overline{\mathcal{D}_R^{  k\downarrow}} \hspace{1mm} \epsilon_{ijk}a_4  a_j  a_i  \hspace{.5mm} v  \vspace{3mm}\\

+ \hspace{.5mm} \overline{\mathcal{U}_R^{  k\downarrow}} \hspace{1mm} \epsilon_{ijk}a_5  a_j  a_i  \hspace{.5mm} v  +  {\mathcal{U}_L^{  i\uparrow}} \hspace{.5mm} a_5  a_4  a_i  \hspace{.5mm} v 
\vspace{3mm}\\

+ \hspace{.5mm} \overline{\mathcal{U}_R^{  k\uparrow} }\hspace{.5mm} \epsilon_{ijk}a_j  a_i  \hspace{.5mm} v +    {\mathcal{U}}_L^{  i\downarrow} \hspace{.5mm} a_4   a_i  \hspace{.5mm} v  \vspace{3mm}\\
 
 +   \hspace{.5mm} {\mathcal{D}}_L^{  i\downarrow} \hspace{.5mm} a_5   a_i  \hspace{.5mm} v  +    \overline{{\mathcal{E}}^{-\uparrow}_R} \hspace{1mm} a_5   a_4  \hspace{.5mm} v \vspace{3mm}\\

+ \hspace{.5mm} \mathcal{D}_L^{i\uparrow} \hspace{.5mm} a_i  \hspace{.5mm} v   + \hspace{.5mm}\overline{\mathcal{E}^{-\downarrow}_R} \hspace{.5mm}a_4  \hspace{.5mm} v     \vspace{3mm}\\

+ \hspace{.5mm} \overline{\mathcal{V}^{\downarrow}_R} \hspace{.5mm}a_5  \hspace{.5mm} v +   \overline{\mathcal{V}^{\uparrow}_R} \hspace{.5mm}\hspace{.5mm} v,
\end{array}
\end{equation}

\noindent where we use the convention that the bar over a particle flips the chirality and $\sigma_z$ spin value.  For example, the anti-particle of the $1/2\hspace{.5mm}\sigma_z = + 1/2$ sterile right-handed neutrino is denoted  $\overline{\mathcal{V}^{\uparrow}_R}$.  It is itself left-handed, and has eigenvalue -1/2 under the spin operator $R_{\sigma_z}:=R_{i\epsilon_3}$.

Rewriting in terms of division algebras, and regrouping, gives 
\begin{equation}\begin{array}{lll}\label{PsiL_div}
\Psi_{\textup{L}}^{\downarrow} + \Psi_{\textup{L}}^{\uparrow} \hspace{1mm}&=&  \hspace{.5mm} \Large(\hspace{.5mm} {\mathcal{V}}_L^{\downarrow} \hspace{.5mm}  i\alpha_1 \alpha_2 \alpha_3 L_{\epsilon_{\uparrow\downarrow}}R_{\epsilon_{\downarrow\downarrow}}  \vspace{3mm}\\

 &+& \mathcal{E}^{-\downarrow}_L \hspace{.5mm} i\alpha_3 \alpha_2 \alpha_1 L_{\epsilon_{\downarrow\downarrow}}R_{\epsilon_{\downarrow\downarrow}}  + \overline{\mathcal{D}^{ k\uparrow}_R} \hspace{.5mm}\epsilon_{ijk}\alpha_j \alpha_i L_{\epsilon_{\uparrow\downarrow}}R_{\epsilon_{\downarrow\downarrow}}   \vspace{3mm}\\

 &+& \hspace{1mm} \overline{\mathcal{U}_R^{  k\uparrow} }\hspace{.5mm} \epsilon_{ijk}\alpha_j\alpha_iL_{\epsilon_{\downarrow\downarrow}}R_{\epsilon_{\downarrow\downarrow}} -    {\mathcal{U}}_L^{  i\downarrow} \hspace{.5mm}  i\alpha_i  L_{\epsilon_{\uparrow\downarrow}}R_{\epsilon_{\downarrow\downarrow}}     \vspace{3mm}\\
 
&+&   {\mathcal{D}}_L^{  i\downarrow} \hspace{.5mm} i\alpha_i  L_{\epsilon_{\downarrow\downarrow}}R_{\epsilon_{\downarrow\downarrow}} +    \overline{{\mathcal{E}}^{-\uparrow}_R} \hspace{1mm} L_{\epsilon_{\uparrow\downarrow}}R_{\epsilon_{\downarrow\downarrow}} \vspace{3mm}\\

&+& \hspace{1mm}  \overline{\mathcal{V}_R^{\uparrow}} \hspace{.5mm}\hspace{.5mm} L_{\epsilon_{\downarrow\downarrow}}R_{\epsilon_{\downarrow\downarrow}},\vspace{3mm}\\

&+& \hspace{1mm}  {\mathcal{V}}_L^{\uparrow} \hspace{.5mm}  i\alpha_3 \alpha_2 \alpha_1 L_{\epsilon_{\uparrow\downarrow}}R_{\epsilon_{\downarrow\uparrow}}  \vspace{3mm}\\

 &+& \mathcal{E}^{-\uparrow}_L \hspace{.5mm} i\alpha_3 \alpha_2 \alpha_1 L_{\epsilon_{\downarrow\downarrow}}R_{\epsilon_{\downarrow\uparrow}}  + \overline{\mathcal{D}^{ k\downarrow}_R} \hspace{.5mm}\epsilon_{ijk}\alpha_j \alpha_i L_{\epsilon_{\uparrow\downarrow}}R_{\epsilon_{\downarrow\uparrow}}   \vspace{3mm}\\

 &-& \hspace{1mm} \overline{\mathcal{U}_R^{  k\downarrow} }\hspace{.5mm} \epsilon_{ijk}\alpha_j\alpha_iL_{\epsilon_{\downarrow\downarrow}}R_{\epsilon_{\downarrow\uparrow}} +    {\mathcal{U}}_L^{  i\uparrow} \hspace{.5mm}  i\alpha_i  L_{\epsilon_{\uparrow\downarrow}}R_{\epsilon_{\downarrow\uparrow}}     \vspace{3mm}\\
 
&+&   {\mathcal{D}}_L^{  i\uparrow} \hspace{.5mm} i\alpha_i  L_{\epsilon_{\downarrow\downarrow}}R_{\epsilon_{\downarrow\uparrow}} +    \overline{{\mathcal{E}}^{-\downarrow}_R} \hspace{1mm} L_{\epsilon_{\uparrow\downarrow}}R_{\epsilon_{\downarrow\uparrow}} \vspace{3mm}\\

&-& \hspace{1mm}  \overline{\mathcal{V}_R^{\downarrow}} \hspace{.5mm}\hspace{.5mm} L_{\epsilon_{\downarrow\downarrow}}R_{\epsilon_{\downarrow\uparrow}}\Large)\hspace{1mm}s^*S^*.
\end{array}
\end{equation}

At this point, we see that rewriting our Dirac action $\ell^D_{10}$ as the Weyl action $\ell_{10}$ allows us to evade the fermion doubling problem.  That is, the model successfully incorporates $\mathfrak{sl}(2,\C)$ spacetime symmetries, without doubling the spinor space.  Hence, it passes checkpoint $\langle 2 \rangle$.  
%Solution copied by kk.

Finally, readers may observe above that internal symmetries are described entirely using the left-multiplication algebra, while the local spacetime symmetries are are described using right-multiplication.  This model cleanly separates the two.  Again, this was an unexpected finding.  It also demonstrates that this model conforms to the Coleman-Mandula theorem, and hence passes checkpoint $\langle 1 \rangle$.

%Beth:  Can we describe how we are effectively going from Cl(10) to Cl(9) in this step?  Dynkin diagrams?

%\section{An alternate characterization}

%Instead of spinor-vector conditions, can see the constraint as a requirement that symmetries all act from one side of the spinor.  Good for spinor-helicity formalism where internal symmetries would then be algebraically internal in a very literal sense. 

%\section{Geometrizing}

%Reinterpreting this algebraic model in terms of a geometric one is trivial.  As was demonstrated as early as~\citep{UFTcg}, these algebraic building blocks can be rewritten 

\section{Reintroduction of $\mathfrak{u}(n)$}

It may at some point be useful to note that a $\mathfrak{u}(n)$ Lie subalgebra is easily obtainable by a single constraint on $\mathfrak{so}(2n)$, similar to the multivector condition that gave us $\mathfrak{su}(n)$.  While the multivector condition gave us $\mathfrak{su}(n)$, the alternative constraint
\begin{equation} \begin{array}{c}\label{unconstraint}  [ r_{k_1k_2} \Gamma_{k_1}\Gamma_{k_2}, \hspace{.5mm}\Psi_{\textup{L}}]\hspace{1mm} = r_{k_1k_2}' \Gamma_{k_1}\Gamma_{k_2}\Psi_{\textup{L}} 
\end{array}\end{equation}
\noindent provides us with $\mathfrak{u}(n)$ for some $r_{k_1k_2}'\in\R,$ not necessarily equal to $r_{k_1k_2}.$

%\section{Operator interpretation}

%Throughout much of quantum theory, the canonical transformation rule of an operator $\hat{O}$ under a group $G$ represented by $\rho(G)$ is given by conjugation:
%\begin{equation} \hat{O}' = \rho(G) \hspace{1mm}\hat{O}\hspace{1mm} \rho(G)^{-1}.
%\end{equation}
%\noindent Given that 

%Mention that operators typically also transform under conjugation in quantum theory.  Could replace Spinor-Vector constraint with Spinor-Operator constraint.

%\section{Group theory conjecture}

%The use of simultaneous group actions allowed us in this article to break one group representation into the group representation of one of its subgroups.  This  leads one to wonder:  could it  be the case in general that any subgroup may be identified via simultaneous group actions of a larger group in which it is embedded? 

%Can every subgroup of a group be given by a simultaneous group action constraint of the original group?

%\section{relation of cascade to real structures - Barrett notes p32}

%Is there a new notion of dagger for each step?

%Can you think of each Z2 as a new chirality operator?

% \section{Spinors to vectors}
 
 %We just did vectors to spinors.  What about spinors to vectors?

 \section{Conclusion}
 
The purpose of this second article was to work through the implications of the new model constructed in~\citep{fr1}.  We listed a set of five checkpoints that serve as minimum conditions that an ideal algebraic description of particle physics should pass.   Namely, such an ideal model should 

$\langle 1 \rangle$  Heed or evade the Coleman-Mandula theorem,

$\langle 2 \rangle$  Solve the fermion doubling problem,

$\langle 3 \rangle$  Explain the Standard Model's chirality,

$\langle 4 \rangle$  Exclude B-L gauge symmetry at low energies,

$\langle 5 \rangle$ Explain the existence of three generations.

Then, we explained how~\citep{fr1} excluded B-L gauge symmetry and demonstrated the correct chiral fermion representations, thanks to the multivector condition (checkpoints $\langle 4 \rangle$ and $\langle 3 \rangle$).  
 
 Of   significance was the serendipitous appearance of spacetime symmetries.  Namely, we constructed a Weyl $\mathfrak{spin}(10)$ representation from the $*_{\overleftarrow{\mathbb{H}}}$-invariant part of the known Dirac  $\mathfrak{spin}(10)$ representation.  It so happened that the maximal $\CLten$ subspace commuting with this Weyl $\mathfrak{spin}(10)$ representation produced the correct $\mathfrak{sl}(2,\C)$ spacetime symmetries.  This allowed us to cross checkpoint $\langle 2 \rangle$.
 
 Also fortuitous was the finding that internal symmetries can find their definition using the pure left-multiplication algebra of $\RCHO,$ while spacetime symmetries find their definition using the pure right-multiplication algebra.  That is, internal and spacetime symmetries exhibit a clean separation, \citep{local}, (checkpoint $\langle 1 \rangle$).

  At this point, we see that \citep{fr1} has passed checkpoints $\langle 1 \rangle$, $\langle 2 \rangle$, $\langle 3 \rangle$, $\langle 4 \rangle$.  We do not attempt here to solve the three-generation problem (checkpoint $\langle 5 \rangle$), but such steps are being made elsewhere.  Interested readers are directed to a new line of research, known as \it Bott Periodic Particle Physics, \rm first introduced in~\citep{fock} and \citep{osmu-furey}.  Alternatively, please see~\citep{fr3} and~\citep{dias}.

%We next demonstrate an algebraic constraint related to the multivector condition of~\citep{fr1} that sends $\mathfrak{so}(2n) \mapsto \mathfrak{u}(n)$, instead of  $\mathfrak{so}(2n) \mapsto \mathfrak{su}(n).$  

%This paper closes by tying the five types of algebraic symmetry breaking constraints of~\citep{fr1} into one overarching theme.  Namely, each constraint may be seen to arise as the result of setting two distinct group actions equal on the same fermionic space.  On a more abstract level, we pose the question:  Could every Lie subgroup be identified in an analogous way?

\begin{acknowledgments}  %Old acknowledgements and biblio:  An early version of these results was first presented in a recorded talk for the 16th Marcel Grossmann Meeting on the 5th of July 2021,~\citep{mg16}.  Subsequently, we have presented this work in recrded talks at the  Joint Mathematics Meeting on the 6th of April, 2022, and for the Dublin Institute for Advanced Studies on the 14th of April, 2022.   https://www.youtube.com/watch?v=jHHjJTztzkc

These manuscripts have benefitted from numerous discussions with Beth Romano.  The author is furthermore grateful for feedback and encouragement from John Baez, Suk\d{r}ti Bansal, John Barrett, Latham Boyle,  Hilary Carteret, Mia Hughes, Kaushlendra Kumar, Agostino Patella, Shadi Tahvildar-Zadeh, Carlos Tamarit, Andreas Trautner, and Jorge Zanelli.

This work was graciously supported by the VW Stiftung Freigeist Fellowship, and Humboldt-Universit\"{a}t zu Berlin.

%\vspace{1.5cm}

%\noindent \it I'll find my way, find my way, home. - Sia \rm

\end{acknowledgments}

\medskip


\begin{thebibliography}{11}
\bibitem{local}  This author is sceptical of the notion that spacetime is fundamental, and  chooses to forgo Poincar\'{e} in favour of SL(2,$\C$).
\end{thebibliography}
\end{document}